\documentclass[11pt,showpacs,amsmath,amssymb]{revtex4}
\usepackage{graphicx}% Include figure files
\usepackage{dcolumn}% Align table columns on decimal point
\usepackage{bm}% bold math
\usepackage{dsfont, amssymb,amsmath,amscd,latexsym, amsthm, amsxtra,amsfonts, mathrsfs}

\usepackage[all]{xy}
\begin{document}

\title{{ { Report on ``American Option Pricing and Hedging Strategies"}}}
\author{Jinshan \ Zhang \thanks{zjs02@mails.tsinghua.edu.cn}}
\email{zjs02@mails.tsinghua.edu.cn}
\affiliation{ Department of
Mathematical Science,
\  Tsinghua University,\\
\small Beijing 100084, People's Republic of China }
%\date{}
\maketitle
%\begin{tabular}{l}
%\begin{table}
 %\hline
  \vskip 19pt
\begin{center} {\bf Content }\end{center}
   This paper mainly discusses the American option's hedging strategies
   via binomial model and  the basic idea of pricing and hedging American option.
   Although the essential scheme of hedging is almost the same as European option,
   small differences may arise when simulating the process for American option
   holder has more rights, spelling that the option can be exercised at
   anytime before its maturity. Our method is dynamic-hedging  method.\\
\textsl{%MSC(2000): primary 60H10, 60G17; secondary 34A12, 55P10. \\
Keywords: put; American put; call; hedging strategies; price;
option; binomial model; Black-Scholes model.}
\vskip 19pt
% \hline
%\end{table}
\setcounter{equation}{0}
\section{ Brief Introduction }
The framework of the paper is as follows. In the first section I'll
present a very simple example on how to price American option and
how the hedge can be done at length, which is modified
from\cite{CRR79}. Then the computer simulation would be represented
in section II, following the  ``real possible path" (if the
assumption that the stock price will be two possible value in
binomial model holds), say, the underlying simulating stock price
was based on binomial model. In section III some  useful discussion
some comments on pricing and hedging process will be presented.

\setcounter{equation}{0}
\section {A Simple Example of Hedging American Call}
To see the whole hedging process distinctly, I give the following
simple numerical example, which has adapted from Cox, Rubinstein,
Ross \cite{CRR79}. The example is about the American call hedging
process, however, the process w.r.t. put is completely the same as
this one, except selling short instead of borrowing money to buy
stocks. Thus this naive example could reflect the basic idea. In
fact the hedging process is a replication of the pricing process,
here we use delta hedging maneuver, see reference
\cite{LLRM96}\cite{JH06}\cite{CRR79}\cite{BW98}. If the current
market price $M$ ever differed from its formula (theoretic) value
$C$. If $M> C$, we would hedge, and if $M < C$, ``reverse hedge", to
try to keep profit. Suppose the values of the underlying variables
(dollars) are

$S = 100$,   $n = 3$,   $K = 100$,   $a = -0.5$,   $b = 1.5$,   $r =
0.1$, $M =45$

In this case,  $p = (r - a)/(b - a) = 0.6$.  The relevant values of
the discount factor are

$r^{-1} = 0.909$,     $r^{-2} = 0.826$,     $r^{-3} = 0.751$

The paths the stock price may follow and their corresponding
probabilities (using probability $p$) are, when $n = 3$, with $S =
100$,

\[
\begin{CD}
100 @>>> 150(.6) @>>> 225(.36) @>>> 337.5(.216)\\
@VVV @VVV @VVV \\
50(.4) @>>> 75(.48) @>>> 112.5(.432)\\
@VVV @VVV \\
25(.16) @>>> 37.5(.288) \\
@VVV\\
12.5(.064)
\end{CD}
\]

when  $n = 2$, if  $S = 150$,

\[
\begin{CD}
150 @>>> 225(.6) @>>> 337.5(.36) \\
@VVV @VVV \\
75(.4) @>>> 112.5(.48)\\
@VVV  \\
37.5(.16)
\end{CD}
\]

when  $n = 2$, if  $S = 50$,

\[
\begin{CD}
50 @>>> 75(.6) @>>> 112.5(.36) \\
@VVV @VVV \\
25(.4) @>>> 37.5(.48)\\
@VVV  \\
12.5(.16)
\end{CD}
\]

\subsection{ Pricing the option}
\label{sec:one-soliton-one-negaton} The pricing process of concrete
model depends on the following fact\cite{LLRM96}: Suppose the market
is viable and complete, the price of American option satisfies
nether formula,
 $C_{n-1}=max(Z_{n-1},(1+r)^{-1}E^{*}(C_{n}|$ $\mathcal {P}_{n-1}))$,
 where $C_{n-1}$ denotes option value at step (time)$n$, $Z_{n-1}$ denotes profit option
 holder will get when exercising the option or not, say, $Z_{n}=max(S_{n}-K,0)$(for a call )
 or $Z_{n}=max(K-S_{n},0)$(for a put), $K$ strike price, $S_{n}$
 stock price at step $n$, $\mathcal {P}_{n-1}$ denotes
 the all the information before $n$ , $E^{*}$ denotes expectation under some
 probability measure (risk neutral probability measure).
 Then computable formula, which can be calculated on computer,
  follows directly from the above results:
Let $C_{n}=P(n,S_{n})$ , then $P(N,x)=Z_{n}$ , $N$ is the maturity.
\begin{displaymath}
P(n,x)=max(Z_{n},(1+r)^{-1}f(n+1,x))
\end{displaymath}
with $f(n+1,x)=pP(n+1,x(1+a))+(1-p)P(n+1,x(1+b))$ and
$p=(b-r)/(b-a)$ . From back forward methods, option value at each
time can be obtained without any difficulties. Using the matlab
program based on the above idea. The current value of the call would
be
\begin{displaymath}
C =42.5995.
\end{displaymath}
\subsection{Hedging the option}
\label{sec:one-soliton-one-negaton} A riskless hedge maneuver can be
simply interpreted as follows: Suppose current stock price is $S$.
We form a portfolio containing $\delta$ shares of stock and the
riskless dollar amount $B$. This will cost $\delta S + B$. At the
end of the period, the value $C$ of this portfolio will be

$P(C=\delta(1+a)S + rB)=p$  denote this value $C$ by $C_{a}$ ;

$P(C=\delta(1+b)S + rB)=1-p$  denote this value $C$ by $C_{b}$ .\\
Since we can select  $\delta$   and $B$  in any way we wish, suppose
we choose them to equate the end-of-period values of the portfolio
and the call for each possible outcome.  This requires that
\begin{displaymath}
 \delta(1+a)S + rB = C_{a}
\end{displaymath}
\begin{displaymath}
 \delta(1+b)S + rB = C_{b}
\end{displaymath}
Solving these equations, we have
\begin{displaymath}
\delta=\frac{C_{b}-C_{a}}{(b-a)S},B=\frac{(1+a)C_{b}-(1+b)C_{a}}{(a-b)S}
\end{displaymath} Using the above formula of $\delta$  and matlab
program, we have the following tree diagram giving the paths the
call value may follow and the corresponding values of $\delta$ :

\[
\begin{CD}
42.5995(.719) @<<< 75.6198(.848) @<<< 134.0909(1.00) @<<< 273.5\\
@AAA @AAA @AAA \\
3.7190(.136) @<<< 6.8182(.167) @<<< 12.5\\
@AAA @AAA \\
0(.00) @<<< 0 \\
@AAA\\
0
\end{CD}
\]

With this preliminary analysis, we are prepared to use the formula
to take advantage of mispricing in the market. Suppose that when $n
= 3$, the market price of the call is $36$.  Our formula tells us
the call should be worth $42.5995$.  The option is overpriced, so we
could plan to sell it and assure ourselves of a profit equal to the
mispricing difference.  Here are the steps we could take for a
typical path the stock might follow.

\textbf{Step 1} ($n = 3$):  Sell the call for $36$.  Take 34.065 of
this and invest it in a portfolio containing  $\delta  = 0.719$
shares of stock by borrowing $0.719(100) -42.5995 = 29.3005$.  Take
the remainder, $45 -42.5995 = 2.4005$, and put it in the bank.

\textbf{Step 2} ($n = 2$):  Suppose the stock goes to $150$ so that
$\delta$ the new is $0.848$.  Buy $0.848 - 0.719 = 0.129$ more
shares of stock at $150$ per share for a total expenditure of
$19.350$. Borrow to pay the bill. With an interest rate of $0.1$, we
already owe $29.3005 (1.1) = 32.2306$.  Thus, our total current
indebtedness is $32.2306 + 19.350 = 51.5806$.

\textbf{Step 3} ($n = 1$):  Suppose the stock price now goes to
$75$. The $\delta$ new is $0.167$.  Sell $0.848 - 0.167 = 0.681$
shares at $75$ per share, taking in $0.681(75) = 51.0750$.  Use this
to pay back part of our borrowing. Since we now owe $51.5806 (1.1) =
56.7387$, the repayment will reduce this to $56.7387 -51.0750 =
5.6637$.

\textbf{Step 4.1} ($n = 0$):  Suppose the stock price now goes to
$37.5$. The call we sold has expired worthless.  We own $0.167$
shares of stock selling at $37.5$ per share, for a total value of
$0.167(37.5) = 6.2625$.  Sell the stock and repay the $5.6637 (1.1)
= 6.2301$ that we now owe on the borrowing without considering the
computing error ($6.2625-6.2301=0.0324$). (In fact, such a error can
be eliminated with high precision, however in the simulation, error
can be cumulated to a little large if the step $N$ goes to infinity
while the precision of the computer is fixed. Fortunately, it has
limit relating to the precision of the computer, this can be seen in
Section II from the hedging performance table). Go back to the bank
and take our initial deposit, which has now grown to $2.4005
(1.1)^{3} = 3.1951$.

\textbf{Step 4.2} ($n = 0$):  Suppose, instead, the stock price goes
to $112.5$. The call we sold is in the money at the expiration date.
Buy back the call, or buy one share of stock and let it be
exercised, incurring a loss of $112.5 - 100 = 12.5$ either way.
Borrow to cover this, bringing our current indebtedness to $6.2301 +
12.5 = 18.7301$. We own $0.167$ shares of stock selling at $90$ per
share, for a total value of $0.167(112.50) = 18.7875$.  Sell the
stock and repay the borrowing without considering the computing
error.  Go back to the bank and take our initial deposit, which has
now grown to $2.4005 (1.1)^{3} = 3.1951$.

\subsection{Remark:}

\textbf{1.} In the above hedging process, we don't care about the
trends of stock price whether it goes up or down. Of course, if the
stock comes into line we can do best thing for us without any loss.

\textbf{2.} If at any step the real price of option equals its
theoretic value, we can buy the option back without concerning of
keeping the portfolio adjusted.

\textbf{3.} In conducting option, we assume every man is rational
(which is an essential assumption of our simulation.). If the man
behaves foolishly and exercises the option at a wrong time, no mater
when he/she exercises the option, (for instance, exercising it as
soon as possible or until the expiration without carrying out it at
optimal time), the value of our portfolio by the above hedging way
would always no less than $S - K$, and our simulation will
illustrate this in the following section. see
\cite{LLRM96}\cite{CRR79}.

\textbf{4.} Instead, we could have made the adjustments by keeping
the number of shares of stock constant and buying or selling calls
and bonds. However, this could be dangerous since the real option
price maybe more than its theoretic value, which next remark
mentions. In large, \textsl{``To adjust a hedged position, never buy
an overpriced option or sell an underpriced option."}\cite{CRR79}

\textbf{5.} The foregoing method is called dynamic hedging
strategies(another name ``hedge-and-forget scheme") while there
exists static hedging strategies(also called ``stop-loss
strategies"). Simulation proves dynamic scheme is better than static
one when applied to European call. Then we can also intuitively
expect the same conclusion when employed in American option since
its special case is European call. The reason why previous
conclusion holds still stands when applied to American option. See
\cite{JH06}\cite{BW98}

\section {Simulation of Hedging Option Process}
\subsection{Assumption}
\textbf{1.} Primary assumption of binomial model.

 \textbf{2.} The
option holder behaves rationally, or equivalently, the option would
be exercised as soon as the optimal time comes (for a call $\delta
=1$, or $S-K \geq$Call Price; for a put  $\delta =-1$, or $K-S
\geq$Put Price).

\textbf{3.} There're no transactions costs.

\textbf{4.} Short is permitted and the rate of money borrowed is
riskless rate.
\subsection{Simulation}
Suppose the underlying variables are as follows (the following data
is selected from \cite{JH06}\cite{BW98}, they utilize Black-Scholes
model to price and hedge European calls, different from our way
based on binomial
model.): \\
Current stock price $S_{0}=\$49$, \\
Strike price $K=\$50$,\\
Stock volatility $\sigma=20\%$ per annum\\
Riskless instantaneous interest rate $R=5\%$per annum\\
Option time to maturity $M=20$ weeks ($0.3836$year)\\
Stock expected return  $\mu=13\%$ per annum \\
The real market option price is $3$.\\
 $N$ denotes the steps of
binomial model (left to be determined); \\
As we know, if we set
\begin{displaymath}
r=RT/N,
1+a=exp(-\sigma\sqrt{T/N}),1+b=exp(\sigma\sqrt{T/N}),p=(b-r)/(b-a).
\end{displaymath}
The price of option will converge to the solution from Black-Scholes
model as $N$ goes to infinity.

\subsection{Simulation Plot:}
The \textbf{yellow} point denotes the changes of stock price at
every step(SP).

The \textbf{green} line denotes the total cashes you have or owed
without considering the stock value you holds at every step (TC).

The \textbf{blue} line denotes the total value of your portfolio
containing stock and bonds at every step (TV).

The \textbf{red} line denotes the total money you'll receive when
the option holder exercises the option or loses the option at every
step before optimal time (TM).

For each $N$, there're four plots including two call-hedge plots and
two put-hedge plots respectively. (all the values in the plots are
per share), see \textbf{APPENDIX}.

From the charts, we could see that the exercising time happens at
maturity or before it, and hedging scheme works very well.
\subsection{Hedging performance:}
If we fixed $N=500$, reiterate the simulation $1000$ times (see
program "gain.m"). The average gain per share of all the gains of
each simulation is $0.6307$, variance is $0.00063542$. Therefore we
know the performance of our hedge can be measured by this variance,
see \cite{JH06}. Performance of dynamic hedging strategies is shown
in the table I. From the table, we can see such a hedging scheme
performs very well for the critical ratio goes to zeros as $N$ goes
to infinity.

\begin{table}
\caption{\label{tab:tablecom} Performance of our dynamic hedging
(The performance measure is the ratio of the variance of the cost of
writing the option and hedging it to the theoretical price of the
option)}
\begin{ruledtabular}
\begin{tabular}{c|c|c|c|c|c|c|c}
%& \multicolumn{2}{c|}{Ras} & \multicolumn{2}{c|}{Liu} &
%\multicolumn{2}{c|}{B-S}
%\\
N steps &  5 & 10 & 25 & 50 & 100 & 250 & 500
\\
\hline step length(weeks) & 4 & 2  & 0.8 & 0.4 & 0.2 & 0.008& 0.004
\\
\hline Hedging performance &  0.0042 & 0.0036 & 0.006 & 0.0031 &
0.002 & 0.001 & 0.0007
\\
\hline average gain(per share) &  0.5571  & 0.6094 & 0.6719 & 0.638&
0.6433 & 0.6338 & 0.631
\end{tabular}
\end{ruledtabular}
\end{table}

\section {Discussion and Remarks}

\textbf{Remark1.} Our simulation shows the dynamic hedging
strategies work perfectly if we accept the assumptions. Hedging
performance reveals its advantages in hedging American call or put.
The program can immediately be applied to simulate the hedging
strategies of European call or put with a small modification. \\

\textbf{Remark2.} In our simulation we have assumed the volatility
is constant with time. In fact in short term, this assumption can be
seen proper and reasonable, however, in a long term (e.g. hedge a
call or put in a long period such as more than one year), errors
would probably raised highly, has $sigma$  been constant over the
periods. In practice, volatility can be calculated, using the past
information of stock price and itself, with statistical methods.
Considering its importance, some classic model has been established
like $ARCH$ and $GARCH$ almost accompany with the advent of
Black-Scholes model, see \cite{En82}\cite{Bo86}.\\

\textbf{Remark3.} As pointed out in Section II, the solution of
binomial model converges to solution of continuous model based on
Black-Scholes model. The key point to hedge the option is replicate
the pricing process, thus pricing the option becomes more essential.
To illustrate this point of view more clearly, let's review our
hedge scheme: in the whole hedging process, our goal is neutralize
delta $\delta$, which is directly related to option price. In this
concrete model(binomial model), call price and put price can be
easily worked out, which is not the same case for American put in
continuous model(Black-Scholes model) since it does not have
explicit solution formula see \cite{LLRM96} chapter 5. Therefore,
many algorithms have been developed to tackle this problem, which
may be summarized to two major approaches. The first approach is a
solution to the integral equation, where the option value is written
as the expected value, under the risk neutral probability measure,
of the option payoffs. Representative algorithms are Binomial method
(the method we used in this project), see \cite{CRR79}; G-J(Geske
and Johnson)method, see \cite{GJ84}\cite{BJ92}; Accelerated Binomial
method, see \cite{Bre91}. The second approach is to directly solve
the Black-Scholes (1973) partial differential equation see
\cite{LLRM96} chapter 5, subject to the boundary conditions imposed
by the possibility of early exercise. Representative algorithms are
Finite Difference \cite{Cour82} and
Recursive method\cite{JSY96}.\\

\textbf{Remark4.} In practice, delta (the changing rate of option
price w.r.t. stock price) is one of the Greek letters taken into
consideration, and other letters such as theta (the changing rate of
option price w.r.t. time), gamma (the changing rate of delta w.r.t.
time or the second derivatives of option price w.r.t. stock price),
vega (the derivative of option price w.r.t. volatility), rho (the
derivative of option price w.r.t. riskless interest rate). The
following words are quoted from \cite{JH06} ``In ideal world, trader
working for financial institutions would be able to rebalance their
portfolios very frequently in order to maintain a zero delta, a zero
vega, and so on. In practice, this is not impossible. When managing
a large portfolio dependent on a single underlying asset, traders
usually zero out delta at least once a day by trading the underlying
asset. Unfortunately, a zero gamma and a zero vega are less easy to
achieve because it is difficult to find options or other nonlinear
derivatives that can be traded in the volume required at competitive
prices (my understanding is that the real market price of option
doesn't equal to theoretic one. As Cox, Ross and Rubinstein say `To
adjust a hedged position, never buy an overpriced option or sell an
underpriced option.'  ). In most case, gamma and vega are monitored.
When they get too large in a positive or negative direction, either
corrective action is taken or trading is curtailed."\\

 {\bf Acknowledgements.} I would like to thank Yves Le Jan for his
  careful reading and helpful comments on the manuscript.

\begin{center}{
\textbf{APPENDIX}}
\end{center}
\newpage
%\section{Appendix}

%\caption{\label{fig:com} }
\begin{center}
\includegraphics[scale=0.92]{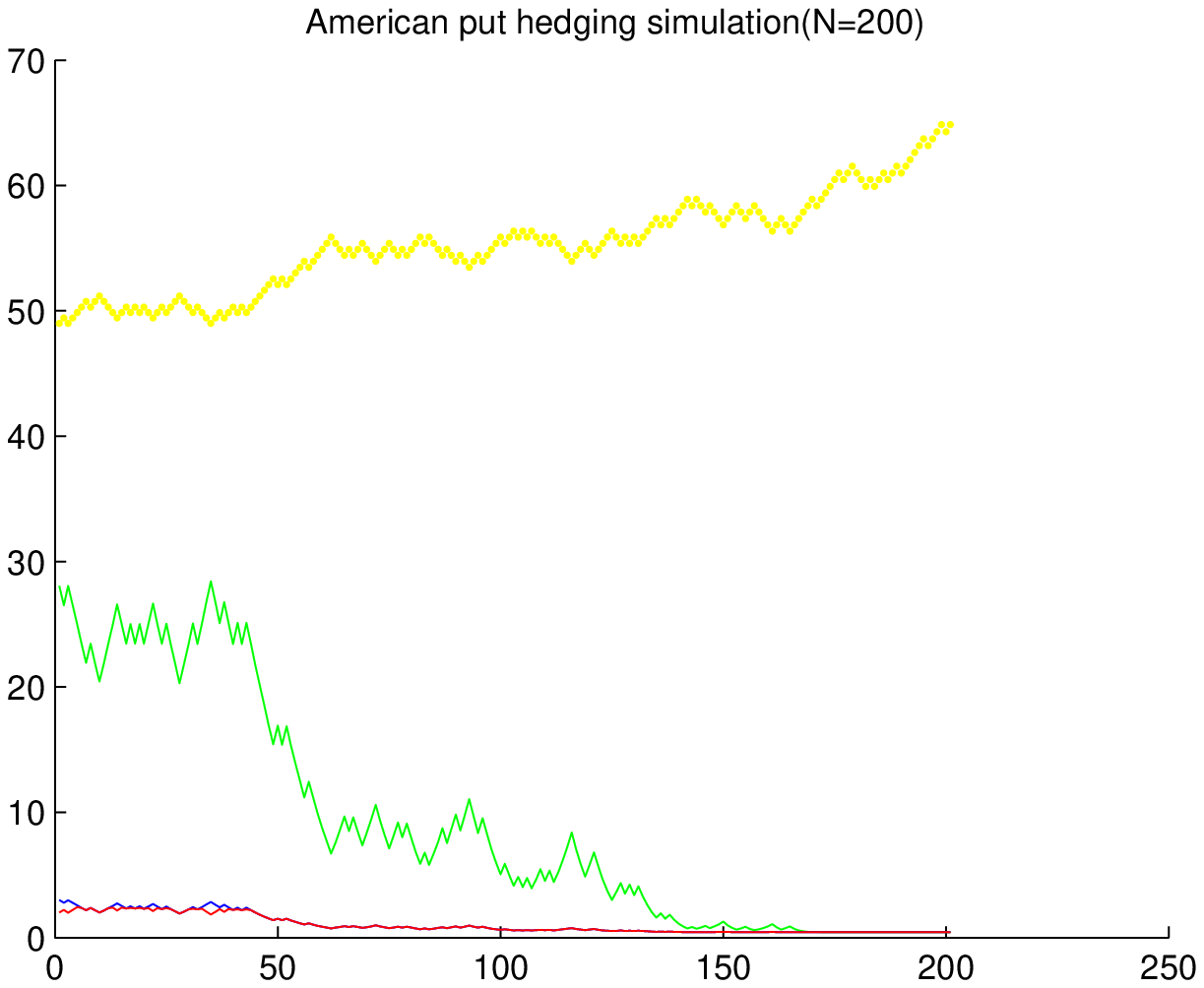}
\end{center}

\begin{center}
\includegraphics[scale=0.92]{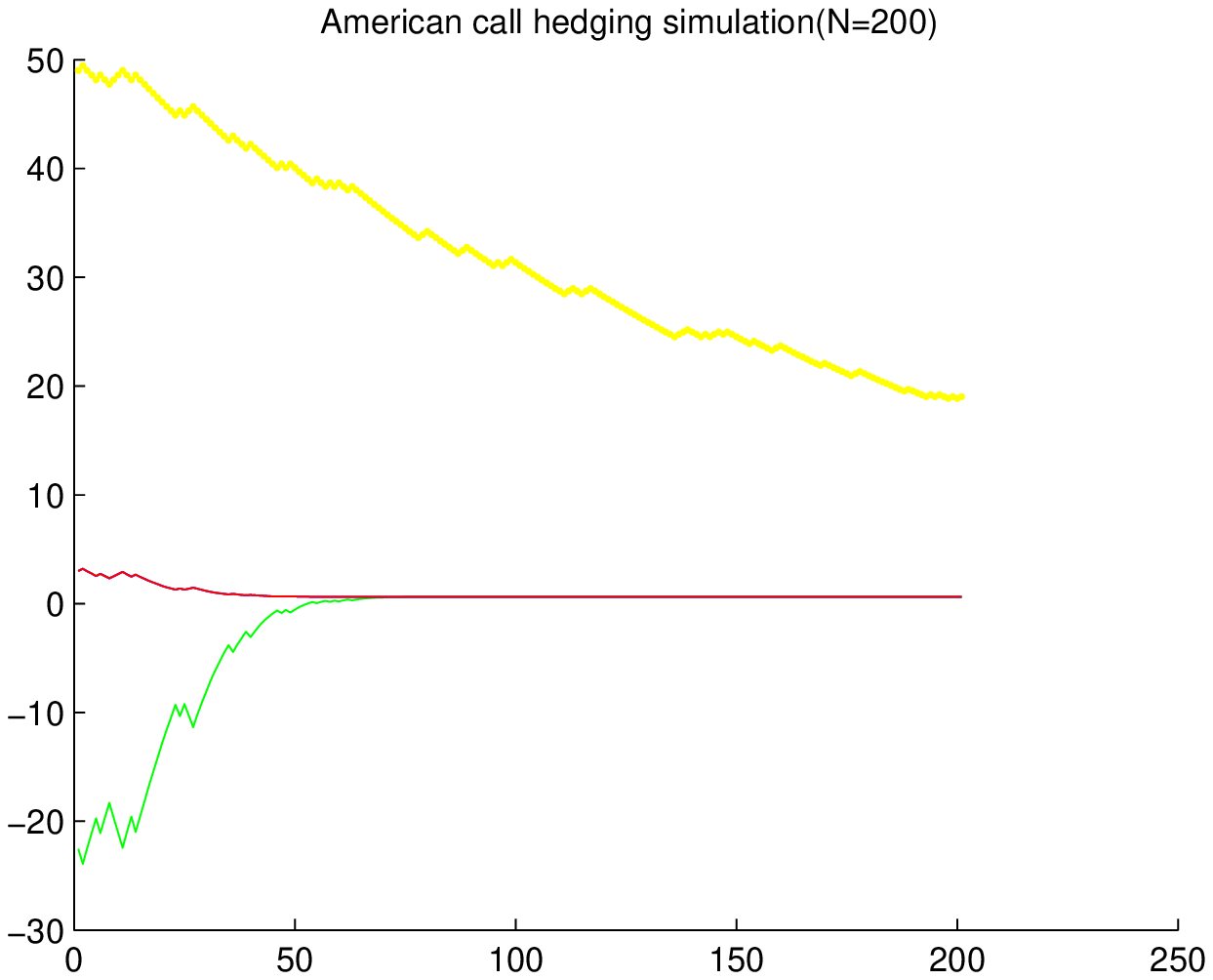}
\end{center}

\begin{center}
\includegraphics[scale=0.92]{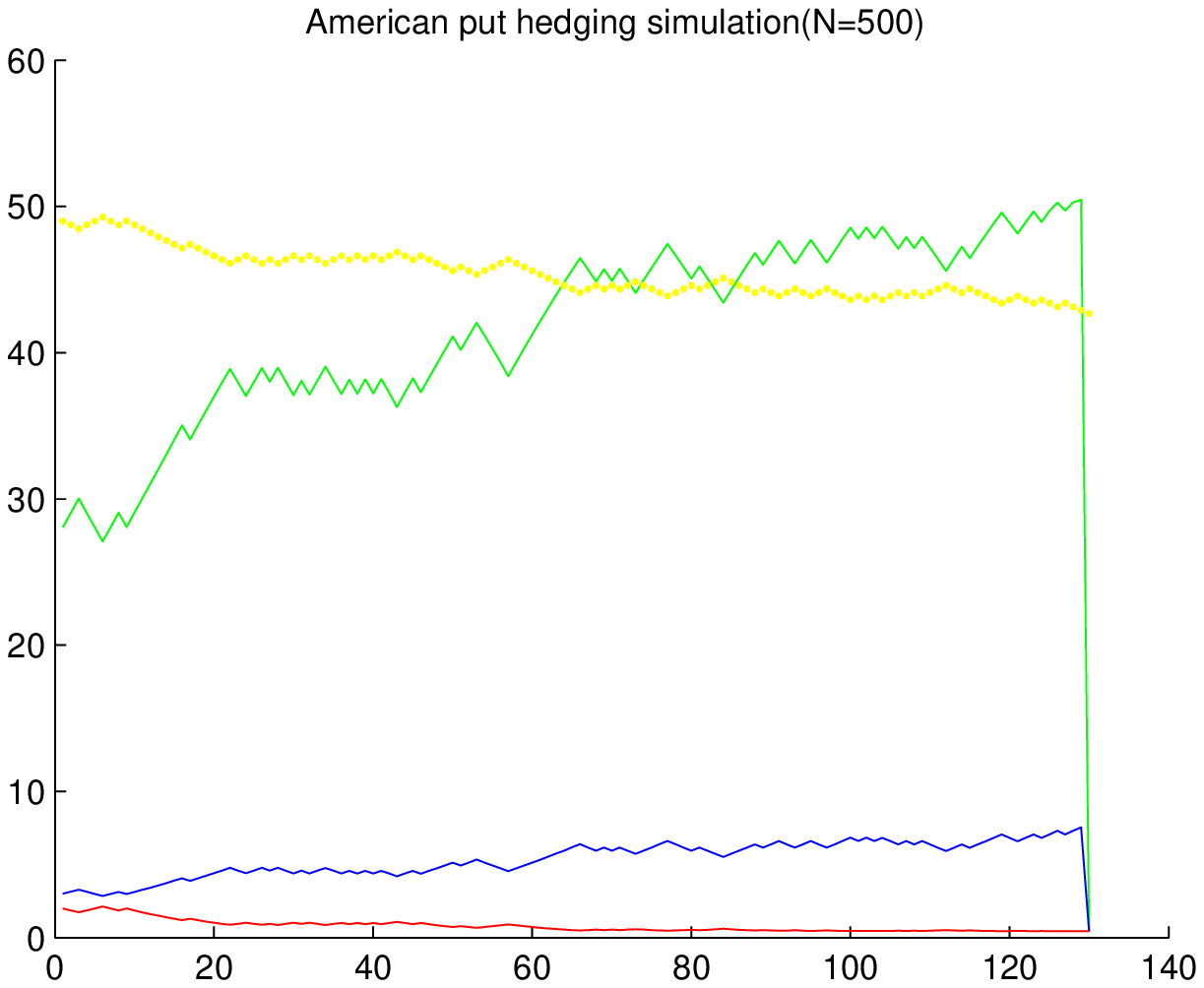}
\end{center}

\begin{center}
\includegraphics[scale=0.92]{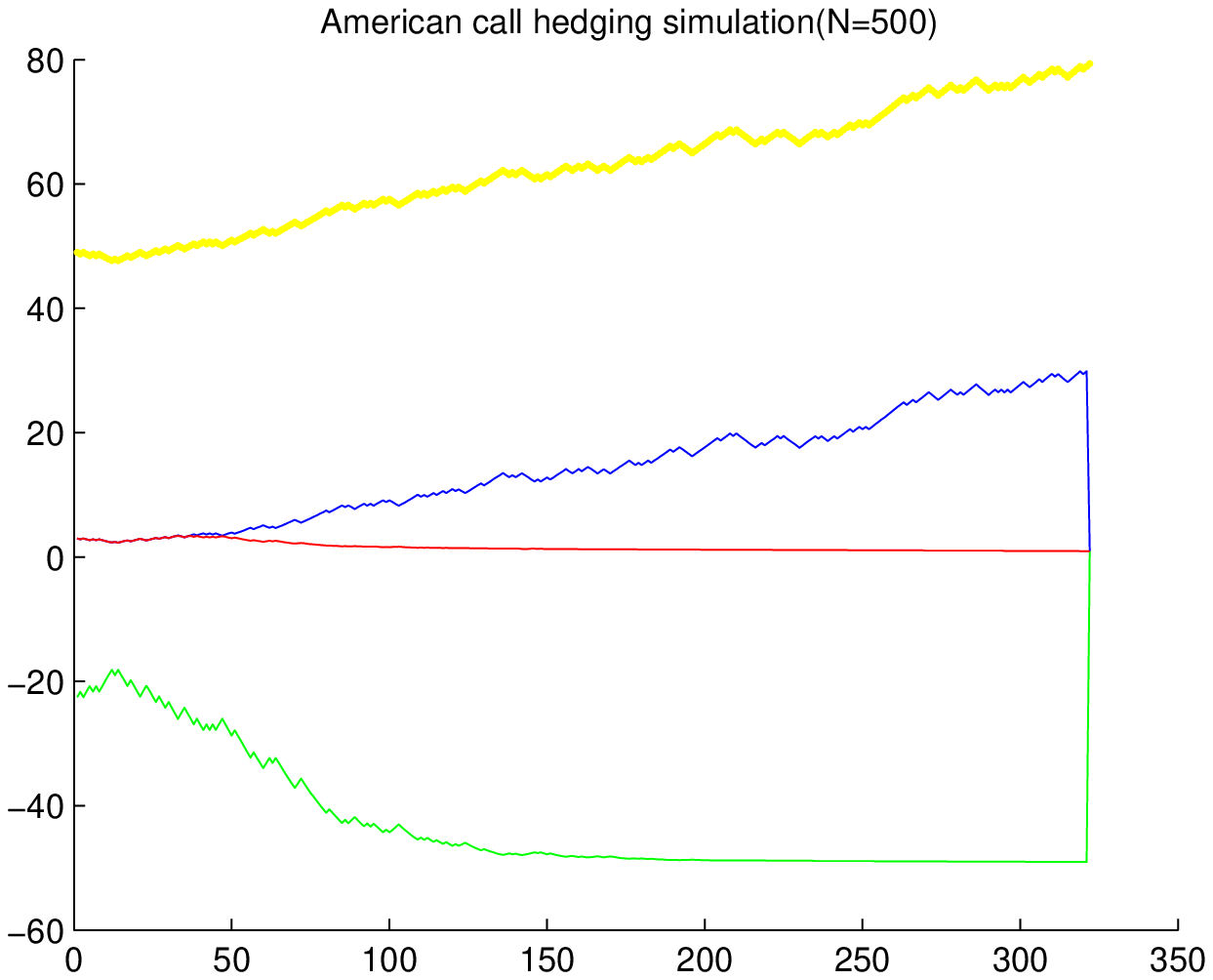}
\end{center}

\begin{center}
\includegraphics[scale=0.92]{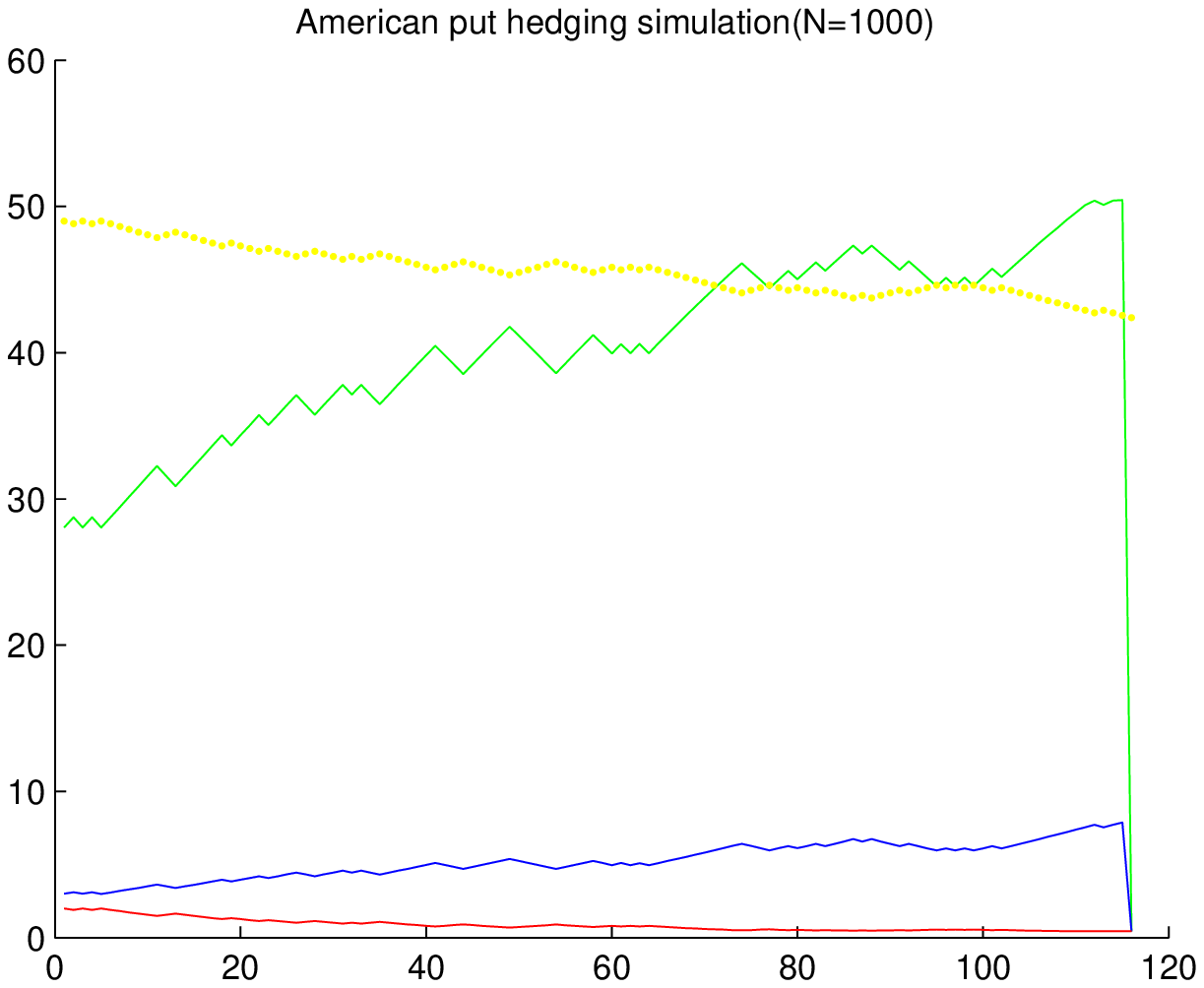}
\end{center}

\begin{center}
\includegraphics[scale=0.92]{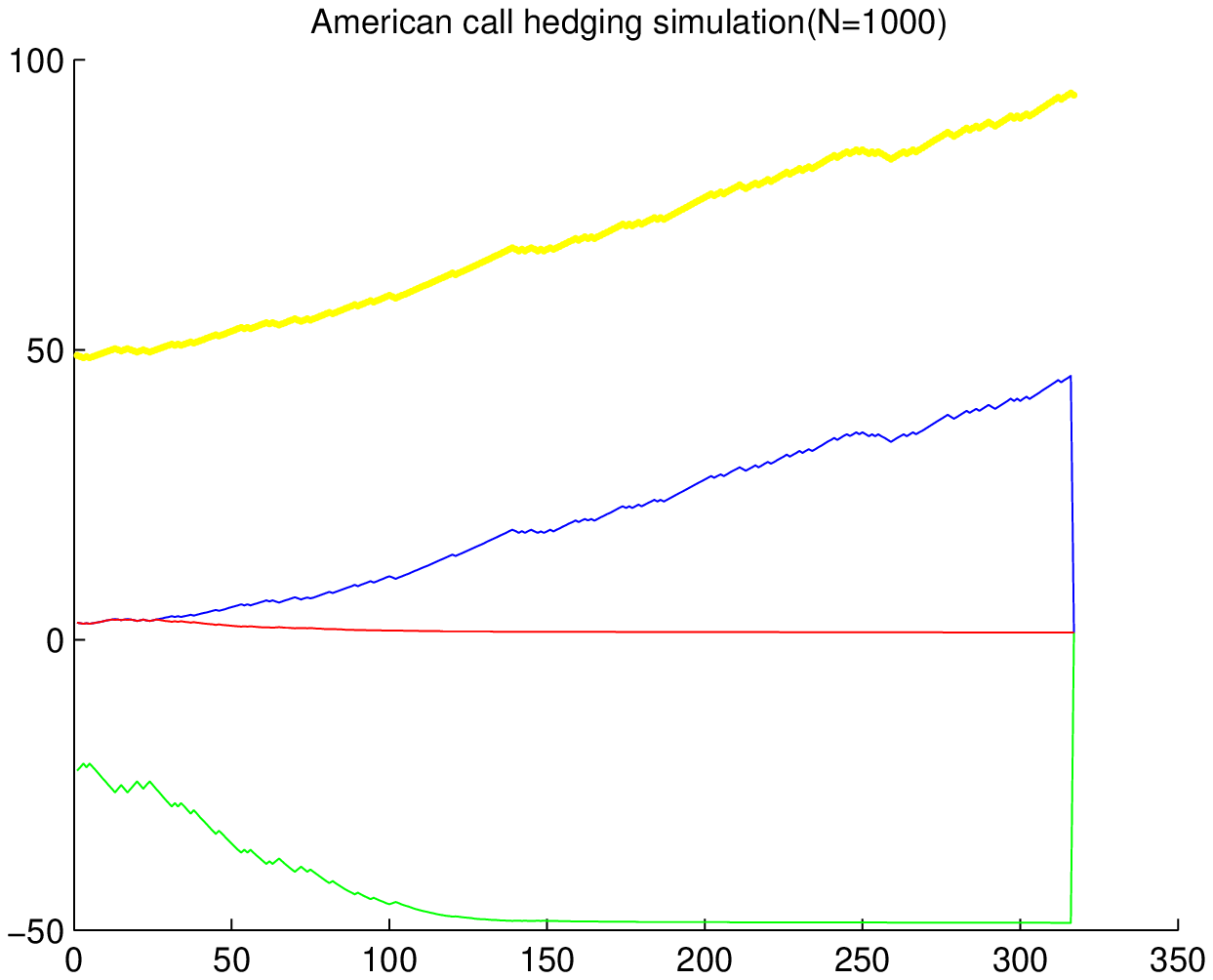}
\end{center}

\end{document}